\documentclass[sigconf]{acmart}

\renewcommand\footnotetextcopyrightpermission[1]{} 
\usepackage{mathptmx} 
\usepackage{booktabs} 
\renewcommand{\textrightarrow}{$\rightarrow$}

\usepackage{hyperref}
\usepackage{amsmath}
\usepackage[caption=false,font=footnotesize]{subfig}
\usepackage{listings}

\newcommand{\mypar}[1]{{\bf #1.}}
\newcommand{\ceil}[1]{\left\lceil#1\right\rceil}
\newcommand{\ceilfrac}[2]{\ceil{\frac{#1}{#2}}}

\def\CC{{C\nolinebreak[4]\hspace{-.05em}\raisebox{.4ex}{\tiny\bf ++}}}

\title[Maximizing CNN Accelerator Efficiency Through Resource Partitioning]{Maximizing CNN Accelerator Efficiency\\
Through Resource Partitioning}
\author{Yongming Shen}
\affiliation{Stony Brook University}
\email{yoshen@cs.stonybrook.edu}
\author{Michael Ferdman}
\affiliation{Stony Brook University}
\email{mferdman@cs.stonybrook.edu}
\author{Peter Milder}
\affiliation{Stony Brook University}
\email{peter.milder@stonybrook.edu}

\renewcommand{\shortauthors}{Y. Shen et al.}

\begin{document}

%
%



\begin{abstract}

Convolutional neural networks (CNNs) are revolutionizing machine learning, but they present significant computational challenges.
Recently, many FPGA-based accelerators have been proposed to improve the performance and efficiency of CNNs.
Current approaches construct a single processor that computes the CNN layers one at a time; the processor is optimized to maximize the throughput at which the collection of layers is computed.
However, this approach leads to inefficient designs because the same processor structure is used to compute CNN layers of radically varying dimensions.

We present a new CNN accelerator paradigm and an accompanying automated design methodology that partitions the available FPGA resources into multiple processors, each of which is tailored for a different subset of the CNN convolutional layers.
Using the same FPGA resources as a single large processor, multiple smaller specialized processors increase computational efficiency and lead to a higher overall throughput.
Our design methodology achieves 3.8x higher throughput than the state-of-the-art approach on evaluating the popular AlexNet CNN on a Xilinx Virtex-7 FPGA. For the more recent SqueezeNet and GoogLeNet, the speedups are 2.2x and 2.0x.
\end{abstract}

\maketitle

\thispagestyle{empty}

\section{Introduction}
The rapid adoption of convolutional neural networks (CNNs) has transformed machine learning.  CNNs have been embraced across a wide array of fields, such as recommendation systems~\cite{van2013deep}, natural language processing~\cite{collobert2008unified}, and computer vision~\cite{krizhevsky2012imagenet,simonyan2014very,iandola2016squeezenet,szegedy2015googlenet}.  In particular, image object recognition has become the de facto benchmark for CNNs, with new networks shattering all prior records in object detection and classification every year.

However, improvements in CNN accuracy are accompanied by a rapid increase in computational cost.  CNNs have already grown to the point where multi-core CPUs are no longer a viable computing platform.  At the same time, while GPUs offer adequate performance, GPU power consumption brings its own set of challenges, particularly at data-center scale.  As a result, FPGAs have seen a surge in interest for CNN acceleration due to their programmable, massively parallel, and power-efficient computing substrate.  The combination of high performance and power efficiency on machine learning tasks is leading to the adoption of FPGAs in data center environments~\cite{catapult}.

CNNs comprise multiple computation \emph{layers}, whose inputs are arrays of different dimensions.
The prior state of the art for using FPGAs for CNNs is to implement an accelerator, which we call a \emph{convolutional layer processor} (CLP), that processes the layers iteratively, one by one.
A CLP design is parameterized by the dimensions of its computational grid; its speed depends on the compatibility of these dimensions with the CNN layers it computes.
To achieve peak performance, the CLP parameters are jointly optimized for the ensemble of the layers to maximize the collective throughput of the accelerator.  This approach closely follows from an ASIC accelerator design flow, where a given hardware design is optimized for the ensemble of the benchmarks that will run on it, so as to perform well for all likely workloads that will be used once the ASIC is deployed.

\begin{figure}[t]
  \vspace{15pt}
  \centering
  \includegraphics[width=\columnwidth]{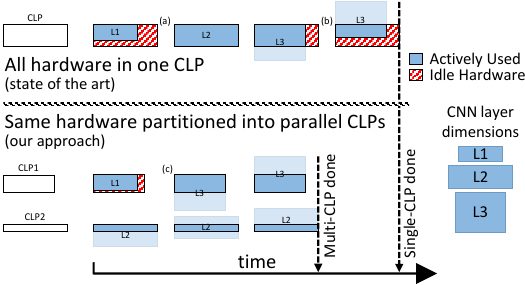}
  \caption{Operation of convolutional layer processors (CLPs) on a three-layer CNN.  Our Multi-CLP approach uses the same total hardware resources as the Single-CLP. However, the Multi-CLP partitioned hardware closely matches the CNN layers, minimizing idle hardware and improving performance.}
  \label{fig:overview}
\end{figure}

We observe that jointly optimizing one CLP for all CNN layers leads to a dynamic underutilization of FPGA resources, giving up performance that could be achieved on the FPGA platform.
Although the CLP is optimized for maximum throughput, the fixed dimensions of the computational grid are sub-optimal for some, or even all, of the individual layers.
Figure~\ref{fig:overview} (top) illustrates this problem.  The Single-CLP hardware (white box) iteratively processes the three layers (blue boxes). The dimensions of the hardware and the layers are represented by the size and shape of the boxes.
L1 is smaller than the CLP dimensions, leaving some hardware unused when computing this layer~(Figure~\ref{fig:overview}(a)). L2's size exactly matches the CLP, but L3's dimensions exceed the CLP size. Therefore, the CLP computational grid must be used iteratively to compute different parts of L3 (first, its top portion, then, its bottom portion), again underutilizing the available hardware (Figure~\ref{fig:overview}(b)). On the popular AlexNet CNN~\cite{krizhevsky2012imagenet}, an ``optimal'' Single-CLP derived from the state-of-the-art methodology~\cite{zhang2015optimizing} has dynamic utilization of less than 24\%. This means that, on average, more than three quarters of the CLP's arithmetic units (multipliers and adders built from the FPGA's DSP slices) remain unused.

To overcome this problem, we propose a new CNN accelerator design that partitions FPGA resources among multiple CLPs, which operate on multiple images concurrently.
We illustrate the operation of Multi-CLP in Figure~\ref{fig:overview} (bottom), where the hardware resources are partitioned among two smaller CLPs that operate in parallel on different images.
Note that the two CLPs are specialized and have different dimensions; this allows CLP1 to work well for L1 and L3, while CLP2's dimensions are compatible with L2. The key is that these sizes allow the layers to be processed with very little idle hardware, enabling the Multi-CLP to do the same amount of work in less time (Figure~\ref{fig:overview}(c)).

We develop an optimization algorithm that, given CNN layer dimensions and a resource budget, computes a partitioning of the FPGA resources into multiple CLPs for an efficient high-performance design.
Our algorithm runs in minutes and produces a set of CLP dimensions.  We then use these dimensions to parameterize a CLP design specified using high-level synthesis (HLS), combining the resulting CLPs to form a complete CNN implementation.

Our results demonstrate that partitioning FPGA resources into multiple CLPs can achieve over $90$\% arithmetic unit utilization, in some cases close to 100\%. Our design methodology achieves 3.8x higher throughput than the state-of-the-art approach for the popular AlexNet CNN on a Xilinx Virtex-7 FPGA. For the more recent SqueezeNet and GoogLeNet, the speedups are 2.2x and 2.0x.

The rest of the paper is organized as follows.
In Section~\ref{sec:background}, we provide background on CNNs.
Section~\ref{sec:utilization-problem} describes the state-of-the-art FPGA implementation and analyzes the inefficiency of Single-CLP accelerators.
Section~\ref{sec:new-design} presents our Multi-CLP optimization methodology.
Section~\ref{sec:hls} describes our design and implementation and Section~\ref{sec:evaluation} details experimental results.
Section~\ref{sec:related} discusses related work and we conclude in Section~\ref{sec:conclusions}.

\section{CNN Background}
\label{sec:background}

\begin{figure}
\centering
\includegraphics[width=\columnwidth]{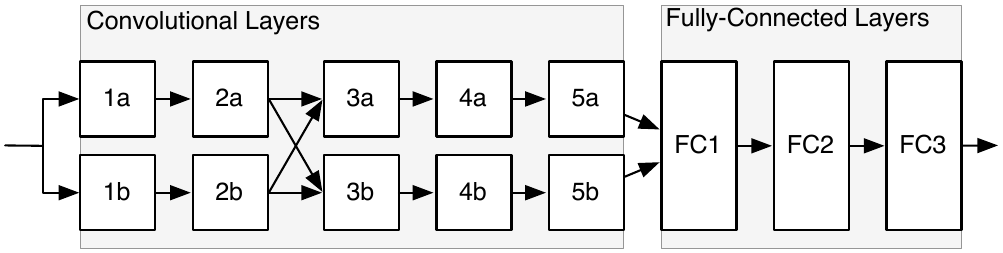}
\caption{AlexNet CNN structure~\cite{krizhevsky2012imagenet}.}
\label{fig:alexnet}
\end{figure}

In typical object recognition examples (e.g.,~\cite{krizhevsky2012imagenet,simonyan2014very}), a CNN passes images through a number of convolutional layers, which convolve the input (an array of two-dimensional matrices called feature maps) with an array of two-dimensional filters, whose values, called weights, were previously learned using an algorithm such as back-propagation.
    Non-linear layers, which typically perform computations such as sub-sampling or activation functions, interleave convolutional layers.
    In the end, the network includes one or more fully-connected layers, each of which performs a number of dot-products across its entire input.
    Figure~\ref{fig:alexnet} shows AlexNet~\cite{krizhevsky2012imagenet}, which contains five stages of paired convolutional layers (e.g., 1a and 1b), followed by three stages of fully-connected layers (FC1--FC3).
    In this figure, the small non-linear layers are omitted.
    As in prior work~\cite{zhang2015optimizing}, we focus on the convolutional layers of the network, because they are the most compute intensive layers.

\begin{figure}
\centering
\includegraphics[width=\columnwidth]{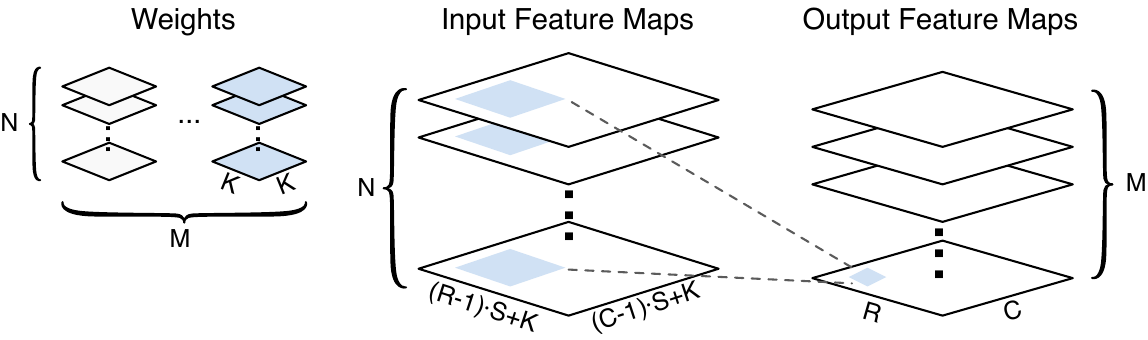}
\caption{Illustration of a convolutional layer.}
\label{fig:conv_layer}
\end{figure}

    Figure~\ref{fig:conv_layer} illustrates a convolutional layer and Listing~\ref{list:nested-loop} presents the pseudo code to compute it. To simplify presentation, we omit biases.
    Each layer takes as input $N$ input feature maps and convolves them with the filters.
    There are $M$ sets of filters; by convolving one set of $N$ filters ($N\times K\times K$ weights) with the input feature maps, one of the $M$ output feature maps is obtained.
    For example, the blue point in the lowest output feature map in Figure~\ref{fig:conv_layer} is computed by taking the dot-product of the blue weights with the portion of the input feature maps shaded in blue.
    All points of the output feature map are computed by
    sliding the blue shaded region around the input feature maps.
    Repeating this process with each of the $M$ sets of filters, we compute each of the $M$ output feature maps.

\section{Resource Utilization Problem}
\label{sec:utilization-problem}
    A common approach for building a CNN accelerator is what we call a \emph{convolutional layer processor} (CLP), which computes the nested loop in Listing~\ref{list:nested-loop}. Because a CNN has multiple convolutional layers, the same CLP is used to process all layers, one by one. Because different layers have different dimensions ($M,N,R,C,K,S$), such a ``one size fits all" approach creates a resource utilization problem, as illustrated in Figure~\ref{fig:overview}. In this section, we analyze how this problem affects a state-of-the-art FPGA CNN accelerator.

\begin{lstlisting}[
float,
mathescape,
caption={Pseudo code of a convolutional layer.},
label=list:nested-loop]
I[N][(R-1)*S+K][(C-1)*S+K] //input maps
O[M][R][C] //output maps
W[M][N][K][K] //weights
for(m=0; m<M; m++)
 for(n=0; n<N; n++)
  for(r=0; r<R; r++)
   for(c=0; c<C; c++)
    for(i=0; i<K; i++)
     for(j=0; j<K; j++)
      wx=W[m][n][i][j]
      ix=I[n][S*r+i][S*c+j]
      O[m][r][c]+=wx*ix
\end{lstlisting}

\begin{figure}
\centering
\includegraphics[width=\columnwidth]{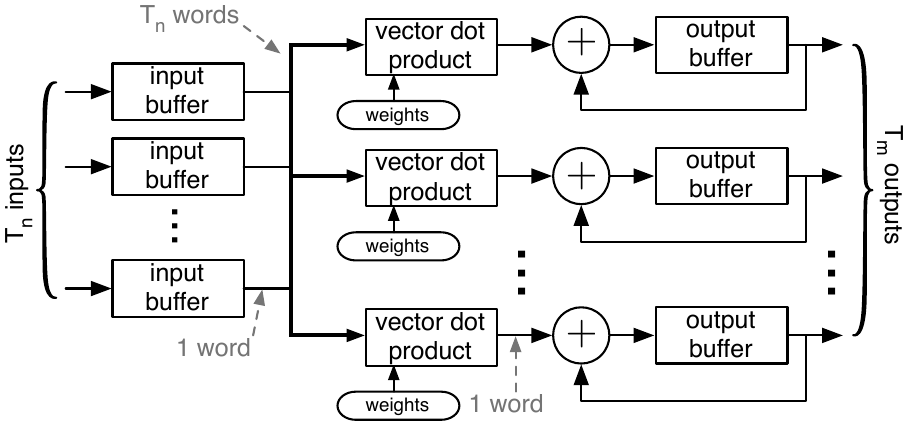}
\caption{CLP design based on~\cite{zhang2015optimizing}. Each dot-product unit takes $T_n$ inputs and $T_n$ weights and produces one output.}
\label{fig:clp}
\end{figure}

\subsection{State of the Art Design}

\label{ssec:clp}
We base our analysis on the design in~\cite{zhang2015optimizing}. This design employs loop transformations, such as loop reordering, tiling, and unrolling to reorder computations and memory accesses, increasing throughput and reducing data transfer. The transformed loop is used as a template for constructing the accelerator.

Using the methodology in~\cite{zhang2015optimizing}, the nested loops in Listing~\ref{list:nested-loop} are transformed into Listing~\ref{list:tiled-loop}, illustrated as a datapath in Figure~\ref{fig:clp}.
The $I_{buf}$, $O_{buf}$, and $W_{buf}$ arrays represent on-chip buffers for input, output, and weight data, respectively.
Copying data in or out of these arrays corresponds to transferring data between the on-chip buffers and off-chip memory.
Double-buffering is used to overlap data transfer with computation and requires provisioning each memory with twice the capacity.
To simplify presentation, Listing~\ref{list:tiled-loop} omits a required boundary check when copying data.

\begin{lstlisting} [
float,
mathescape,
caption={Pseudo code for tiling in a CLP~\cite{zhang2015optimizing}.},
label=list:tiled-loop]
I[N][(R-1)*S+K][(C-1)*S+K] //input maps
O[M][R][C] //output maps
W[M][N][K][K] //weights
Ibuf[Tn][(Tr-1)*S+K][(Tc-1)*S+K]
Obuf[Tm][Tr][Tc]
Wbuf[Tm][Tn][K][K]
for(r=0; r<R; r+=Tr)
 for(c=0; c<C; c+=Tc)
  for(m=0; m<M; m+=Tm) {
   for(n=0; n<N; n+=Tn) {
    irx=r*S:(r+Tr-1)*S+K
    icx=c*S:(c+Tc-1)*S+K
    Ibuf=I[n:n+Tn][irx][icx]
    Wbuf=W[m:m+Tm][n:n+Tn]
    for(i=0; i<K; i++)
     for(j=0; j<K; j++)
      for(tr=0; tr+r<min(R,r+Tr); tr++)
       for(tc=0; tc+c<min(C,c+Tc); tc++)
        for(tm=0; tm<Tm; tm++) #UNROLL
         for(tn=0; tn<Tn; tn++) #UNROLL
          wx=Wbuf[tm][tn][i][j]
          ix=Ibuf[tn][S*tr+i][S*tc+j]
          Obuf[tm][tr][tc]+=wx*ix
   }
   O[m:m+Tm][r:r+Tr][c:c+Tc]=Obuf
  }
\end{lstlisting}

The loops $R$, $C$, $M$, and $N$ are tiled with factors $T_r$, $T_c$, $T_m$, and $T_n$, respectively.
These loop tiling factors control how much data are transferred per buffer refill or write-out, and the order in which data are transferred.
Because the inner-most two loops are unrolled (based on $T_m$ and $T_n$), loop tiling also controls how the compute modules are constructed.
In particular, to implement these two unrolled loops, $T_m$ vector dot-product units are constructed, each of width $T_n$.
An accumulation adder is added after each unit, as shown in Figure~\ref{fig:clp}.
This yields $T_mT_n$ multipliers and adders.

Given a resource budget (e.g.,~a number of DSP slices), one can find the optimal $T_n$ and $T_m$ for a given convolutional layer.
In~\cite{zhang2015optimizing}, a joint optimization is performed to create a single CLP to compute \emph{all} of the convolutional layers in the CNN.
The optimization finds the $(T_n,T_m)$ that maximize the aggregate performance of the CLP.

\subsection{Arithmetic Unit Utilization Problem}

Although the methodology in~\cite{zhang2015optimizing} produces a CLP optimized for the collective performance of all convolutional layers, we observe that its speed is limited by the fact that different convolutional layers of the CNN have different dimensions, but all are computed on the same $(T_n,T_m)$ CLP.
Thus, the CLP that gives the best performance \emph{across all layers} is not necessarily well suited for any one layer.
Because the limiting factor of performance is the number of parallel arithmetic units in the CLP, the cost of the mismatch can be quantified by considering the utilization of the arithmetic units.
That is, we can quantify the percentage of the time that the arithmetic units in the CLP are doing work versus the percentage of the time they are idle.

The primary cause of the utilization penalty is a mismatch between the tile parameters $(T_n,T_m)$ and their corresponding loop sizes $(N,M)$.
In particular, if $N$ is less than $T_n$ or $M$ is less than $T_m$, then there must be cycles where some of the $T_n\times T_m$ multipliers are not used.
For example, following the methodology in~\cite{zhang2015optimizing}, we generated an accelerator for SqueezeNet~\cite{iandola2016squeezenet} that targets the Virtex-7 690T FPGA and uses single precision floating-point arithmetic. The best $(T_n,T_m)$ we obtained is $(9,64)$. However, the $(N,M)$ of layer one of the network is $(3,64)$, therefore $N < T_n$, leading to an arithmetic unit utilization of 33.3\%. The $(N,M)$ of layer two of the network is $(64, 16)$, so $M < T_m$. To make things worse, for layer two, $N=64$ is not a perfect multiple of $T_n=9$, which is another source of underutilization. Eight iterations are needed for $T_n$ to cover $N$. The first seven iterations cover the first 63 input feature maps, leaving only one for the eighth iteration, during which only $1/9$ of the $T_n$ is used. The compound effect of mismatch on both $T_n$ and $T_m$ leads to a utilization of only 22.2\% for layer 2. Overall, analyzing all convolutional layers in SqueezeNet gives an arithmetic unit utilization of 76.4\%.

When fixed-point arithmetic is used, more adders and multipliers can be built using the same DSP slice budget, exacerbating the mismatch problem. The worst case we observed is running AlexNet on a Virtex-7 690T FPGA with 16-bit fixed-point arithmetic units, which has an overall arithmetic unit utilization of less than 24\%.

\section{Multi-CLP Design}
\label{sec:new-design}
    To improve the resource utilization and thus CNN performance, we propose Multi-CLP accelerators, where the available resources are partitioned across several smaller convolutional layer processors rather than a single large one.
    The advantage comes from the CLPs having different sizes, more closely matching the dimensions of the CNN layers.
    This approach is possible because CNN accelerators process many input images, allowing CLPs to concurrently work on independent inputs.

    To construct a Multi-CLP accelerator for a given resource budget and CNN structure, one must decide how many CLPs to use, how to partition the resources among them, and how to distribute and schedule the processing of individual convolutional layers from multiple images on the CLPs.
    We describe (a) the operation of a Multi-CLP, (b) a model to predict CLP costs and performance given its parameters, and (c) an optimization algorithm to find the best Multi-CLP design for a given resource budget and set of CNN layers.

\subsection{Multi-CLP Accelerators for CNNs}
\label{ssec:vpipe}

    Due to the feed-forward nature of the CNN structure, it is natural to think of the layers of the CNN as a pipeline.
    Therefore, one way to construct a CNN accelerator with multiple CLPs is to implement a separate CLP for each layer~\cite{li2016high}.
    An accelerator for an $L$-stage CNN would have $L$ CLPs and would operate on $L$ independent input images.
    (That is, CLP1 would work on image $i$ while CLP2 works on image $i-1$, etc.)
    This would have the benefit of allowing each CLP to be optimized solely for the dimensions of one CNN layer, improving efficiency.

    A limitation of this approach, however, is that it requires the number of CLPs to be equal to the number of convolutional layers.
    This poses several problems for practical CNNs.
    First, it forces the design to divide the on-chip BRAM resources, reducing the buffer size of each CLP.
    As a result, the ability to exploit data reuse in each CLP diminishes, greatly increasing the overall memory bandwidth requirement and slowing down each CLP.
    Second, this one-to-one mapping of CLPs to convolutional layers requires coordinating a large number of accesses to off-chip memory, which is costly in terms of performance and logic resources.
    Third, each CLP has an overhead cost (i.e., control logic for address calculation and loop index state machine). If there are many CLPs, significant resources are devoted to control logic instead of CNN computation.

    To address these problems, we target Multi-CLP designs that minimize the number of CLPs in an accelerator.
    This approach requires at least one CLP in the design to compute multiple CNN layers.
    We use a static assignment of layers to CLPs, where each layer is bound to one CLP.
    Layers assigned to the same CLP need not be adjacent in the CNN structure.

\begin{figure}
\centering
\includegraphics[width=\columnwidth]{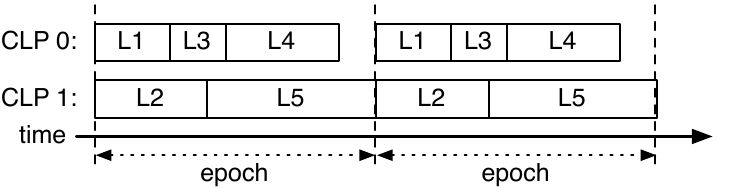}
\caption{Example schedule for a Multi-CLP system.}
\label{fig:clp_pipe}
\end{figure}

    The timeline of the accelerator operation is divided into \emph{epochs}. In each epoch, each CLP sequentially processes its layers, with each layer having its own independent data.
    The epoch ends when all CLPs finish.
    Figure~\ref{fig:clp_pipe} shows an example where CLP0 processes three layers (L1, L3, and L4) and CLP1 processes two layers (L2 and L5).
    In each epoch, each CLP only consumes data generated during the \emph{previous} epoch, avoiding data dependencies within a epoch.
    For example, the output produced by L1 in epoch $i$ will be used as input for L2 in epoch $i+1$.
    This means that processing an image requires five epochs, therefore data from five different images will be in flight at a time.
    Because the intermediate data are typically too large to hold on chip, all CLPs read their inputs from and write their outputs to off-chip memory.

If the evaluation latency must be limited further, one can constrain the layer assignment such that layers for the same CLP are adjacent in the CNN structure. This way, a CLP can process multiple layers for an image in a single epoch, and the total number of in-flight images is equal to the number of CLPs. This means one can reduce latency by limiting the number of CLPs, but this is achieved at the cost of throughput.

    There are several considerations for a Multi-CLP system to achieve high throughput; we use these as the targets of our optimization method (Section~\ref{ssec:optimization}).
    First, the epoch length, and thus the system throughput, is limited by the CLP that takes the longest to complete its assigned work.
    For example, in Figure~\ref{fig:clp_pipe}, CLP0 is idle after it finishes L4, until the next epoch begins.
    Second, the convolutional layers assigned to a CLP should have dimensions compatible with the CLP's dimensions to ensure high arithmetic unit utilization.
    Third, the on-chip memory allocated to each CLP is inversely related to the off-chip bandwidth that it requires; larger CLP buffers reduce off-chip data transfer.

\subsection{Modeling CLP Cost and Performance}
\label{ssec:cost}

To find an efficient Multi-CLP design for a CNN, we first construct a model of the CLP costs (DSP slices, BRAMs, memory bandwidth) and speed (cycles).
A CLP is parameterized by its size $(T_n,T_m)$ and the tiling parameters $(T_r,T_c)$ of each of its layers.
The model also uses the dimensions of the convolutional layers: $M, N, R, C, K$, and $S$ (Section~\ref{sec:background}). From these parameters, one can derive the resource use and performance of a CLP.
Because our CLP is based on~\cite{zhang2015optimizing}, a number of the formulas used in our models appear similar, but include several notable differences.

\mypar{Performance Model}
Assuming a fixed frequency target, the performance of a CLP is dictated by the number of cycles needed to compute each of its layers.
Because arithmetic units may be underutilized, the cycle count cannot be calculated by dividing the amount of work by the number of arithmetic units in the design.
Instead, an exact counting of loop iterations based on Listing~\ref{list:tiled-loop} is needed.
In Listing~\ref{list:tiled-loop}, the inner-most two loops are unrolled, thus they do not contribute to the iteration count.
Along the $R$ dimension, the combination of the outer loop (looping over tiles) and the inner loop (looping over elements in a tile) run for $R$ iterations.
Similarly, there are $C$ iterations along the $C$ dimension.
The remaining four loops have $\ceilfrac{M}{T_m}$, $\ceilfrac{N}{T_n}$, $K$, and $K$ iterations, respectively.
Together, the cycle count needed to compute one layer is
\begin{equation*}
\label{eq:cycles}
Cycles = R \times C \times \ceilfrac{N}{T_n} \times \ceilfrac{M}{T_m} \times K^2.
\end{equation*}
Importantly, if a CLP is found to be memory-bandwidth bound, our optimization procedure determines the cycle count by using the data transfer time instead of the cycle count computed by this formula.

\mypar{Modeling Bandwidth Usage}
We are primarily focused on the peak bandwidth use of a CLP, to estimate how much bandwidth is needed to support the maximum computation speed. When the peak bandwidth is unavailable on the target platform, the model must be able to estimate the throughput of the accelerator, taking into consideration how compute may be blocked by data transfer. This allows design space exploration to find the best-performing design under a bandwidth limitation.

A CLP uses double-buffering to overlap data transfer with computation. Layers are processed back to back, allowing data transfer for one layer to be overlapped with computation for another.
The model calculates the start and end times of data transfer and compute, and takes into account the dependencies between them to determine the cycles required to finish computation in the bandwidth-bound cases.

\mypar{Modeling DSP Slice Usage}
The dominant use of DSP slices is the $T_m$ dot-product units (each of size $T_n$) and $T_m$ accumulator adders (see Figure~\ref{fig:clp}); each CLP contains $T_m \times T_n$ multipliers and adders.
For floating-point arithmetic, each multiplier comprises two DSP slices and each adder comprises three.
For 16-bit fixed-point arithmetic, a single DSP slice provides both an adder and multiplier.
Therefore, the respective DSP counts are
\begin{equation*}
\label{eq:dsp}
\begin{split}
NumDSP_{float} &= (3+2)\times T_n \times T_m,\\
NumDSP_{fixed} &= T_n \times T_m.
\end{split}
\end{equation*}

\mypar{Modeling BRAM Usage}
BRAMs are used to construct the various on-chip buffers. Modeling must account for the banking of these buffers, the number of read/write ports a buffer uses, whether double-buffering is used, the capacity and capabilities of a BRAM, and the size of a word.
Here, we assume the word size to be either 16 or 32 bits, although this can easily be extended.
We report usage in terms of Virtex-7 BRAM-18Kb units, which can store 512 32-bit words and operate in ``Simple Dual-Port mode''~\cite{XilinxMem}, allowing one read and one write concurrently.

The input buffer is organized into $T_n$ banks of size $B_i$, which is provisioned to support the computation of all of the layers on a CLP. When computing a layer, each bank stores
\begin{equation*}
[(T_r - 1) \times S + K] \times [(T_c - 1) \times S + K]
\end{equation*} words.
Because the parameters change from layer to layer, each layer needs a different amount of data, requiring $B_i$ to be large enough to support the most demanding layer.
An input bank must be double-buffered to support the overlapping of computation and data transfer, using one read port and one write port.
With 32-bit words, this buffer is constructed with $2 \cdot \ceilfrac{B_i}{512}$ BRAMs.
However, because a single BRAM already provides a read port and a write port, when $B_i \leq 256$, one BRAM is sufficient to construct a double-buffered input bank.

The weight buffer is similar to the input buffer. There are $T_n \times T_m$ banks.
When computing a layer, each weight bank stores a $K \times K$ filter ($K^2$ words).
Thus, of the layers that a CLP computes, the layer with the largest $K$ determines the size of a weight bank.
Other aspects of the weight buffer are modeled in the same way as the input buffer.

The output buffer is organized into $T_m$ banks. When computing a layer, each bank stores $T_r \times T_c$ words.
The output buffer is provisioned for the most-demanding layer and is double-buffered.
However, to support the accumulation used in the CLP, an output-buffer bank requires at least two BRAMs for double-buffering, because the accumulation buffer requires both a read port and a write port.
The remaining aspects of the output buffer model are the same as the input and weight buffer models.

For all buffers, the number of banks is halved for the 16-bit fixed point data type, because pairs of 16-bit words are packed into 32-bit wide BRAMs. Additionally, if a bank stores only a small number of values (fewer than 10), we do not count them toward the BRAM budget, because small memories are implemented as LUTRAMs.

\subsection{Optimization of Multi-CLP Designs}
\label{ssec:optimization}

We now describe an optimization tool that uses the above model to find the fastest Multi-CLP configuration (for a given CNN) that fits within the specified resource constraints.
Because we are targeting FPGAs, we optimize our accelerator only for a specific target CNN.
However, this optimization can be simultaneously applied to multiple target CNNs to jointly optimize their performance.

The optimization takes as input the CNN layer descriptions and target FPGA resource constraints, and produces the parameters to construct and use a Multi-CLP design.
The optimization result includes the number of CLPs and their dimensions $(T_n,T_m)$.
It also includes the distribution of the CNN layers to the CLPs and, for each layer, the $(T_r,T_c)$ parameters, which dictate how the layer is computed by the CLP (Section~\ref{ssec:clp}).

Given a set of parameters, evaluating the model is simple.
However, there are far too many possibilities to perform an exhaustive search for the fastest design that meets the resource constraints, requiring the use of heuristics during optimization.

The optimization process comprises two steps, iteratively repeating these steps until at least one solution that meets the constraints is discovered.
At the beginning of each iteration, a performance target is set.
If the performance target cannot be met at the end of the iteration, a new iteration is started with a slightly lower target.

In each iteration, the first step ($OptimizeCompute$) focuses on the partitioning of DSP slices.
The output of this step is a collection of partition candidates.
Each candidate is a partial solution that specifies the number of CLPs, the $(T_n,T_m)$ of each, and the assignment of CNN layers to the CLPs.

The challenge of $OptimizeCompute$ arises in assigning the CNN's layers to CLPs.
Because the number of possible assignments is exponential with respect to the number of layers, a complete search of this space is impractical.
We mitigate this problem through the observation that, in the best designs, a CLP is assigned ``similar'' layers.
We first produce an ordered list of the layers based on a heuristic (e.g., compute-to-data ratio for bandwidth-limited accelerators or Euclidean distance between $(N,M)$ pairs for compute-bound accelerators).
Then, when we assign layers to CLPs, we only consider candidates where a CLP computes a set of adjacent layers in this order, allowing our search to prune inefficient solutions where incompatible layers would share a CLP.

The second step ($OptimizeMemory$) of the optimization focuses on the partitioning of BRAM slices.
For each candidate from step one, $OptimizeMemory$ finds the $T_r$ and $T_c$ values to use for each layer that minimize the peak memory bandwidth use.
These parameters in turn determine the buffer sizes of each CLP.
A single candidate from $OptimizeCompute$ can result in multiple solutions in $OptimizeMemory$; we consider all of these solutions in the optimization process.
If all candidates have peak memory bandwidth use higher than the available budget, a new iteration of optimization is needed, implying that the final solution will be bandwidth bound.
When estimating the bandwidth requirements of a design during optimization, we allow computation of some CLPs to be blocked by data transfer.
This potentially sacrifices dynamic utilization of some CLPs in the design, but in some cases results in the highest-performing designs overall, despite including bandwidth-bound CLPs that are idle and waiting for data for a small portion of the overall execution.

Listing~\ref{list:optimize} shows the pseudo code of the optimization procedure, which we implemented in \CC.
We separate each iteration into two steps, $OptimizeCompute$ and $OptimizeMemory$, and we use different algorithms for each.
Both steps use memoization to avoid redundant work.
The search can be further sped up by limiting the maximum number of CLPs to consider.
Our \CC\  implementation can complete an optimization of a Multi-CLP accelerator for the GoogLeNet network in several minutes.

\begin{lstlisting} [
float,
morekeywords={procedure},
mathescape,
caption={Pseudo code for Multi-CLP optimization.},
label=list:optimize]
procedure OptimizeMultiCLP($cnn,N_{dsp},N_{bram},bw$)
  $step = 0.005$; $target = 1.00$
  $Cycles_{min}=MinimumPossibleCycles(cnn, N_{dsp})$
  while($target\neq 0$)
    $X = $OptimizeCompute($cnn,N_{dsp},\frac{Cycles_{min}}{target}$)
    if($X = \emptyset$)
      $target = target - step$
      continue
    else
      $A = $OptimizeMemory($cnn, N_{bram}, bw, X$)
      if($A = \emptyset$)
        continue
      else
        return $A$
\end{lstlisting}

Lastly, we note that the same optimization method can be used for Single-CLP accelerator designs by constraining $OptimizeCompute$ to only consider solutions with one CLP.

\section{Design and Implementation}
\label{sec:hls}

The optimization algorithm (Section~\ref{sec:new-design}) determines the characteristics of an optimized Multi-CLP accelerator, producing parameters for a \CC\ high-level-synthesis (HLS) template.
The template is compiled to synthesizable Verilog using Vivado HLS 2016.3.
Our optimizer works for any set of CNN layers and resource budget, and our template supports computation over arbitrary data types.

\subsection{Convolutional Layer Processor Template}

The accelerator template is constructed based on nine parameters: $T_n$ and $T_m$ (to size the CLP compute module), $M_{max}$, $K_{max}$, $in_{size}$, and $out_{size}$ (to size the on-chip bias, weight, input, and output buffers), and $NP$, $WP$, and $MP$
(to specify the number of AXI stream ports for transferring input, weight, and output data).
Each CLP produced with the HLS tool has an auto-generated AXI4-Lite slave interface to trigger the start of computation and AXI stream interfaces for reading and writing data.
For the Multi-CLP designs, each parameterized CLP template is passed through the HLS tool separately, producing independent IP cores that can be inserted into the top-level system and interconnected using a standard AXI crossbar and AXI DataMovers.
One AXI4 port is used at the start of CLP operation to perform a burst transfer of a 32-byte descriptor containing the arguments for the computation ($R$, $C$, $M$, $N$, $K$, $S$, $T_r$, $T_c$).
After these arguments are retrieved and the derived variables are computed (e.g., $r_{steps}$, $c_{steps}$, $m_{steps}$, $n_{steps}$), the design state machine executes the four nested loops shown in Listing~\ref{list:hls-top-level}.

\begin{lstlisting} [
float,
mathescape,
morekeywords={elif},
caption={Pseudo code for an accelerator template.},
label=list:hls-top-level]
compute():
#pragma ARRAY_PARTITION out(1),bias(1)
#pragma ARRAY_PARTITION in(1),weights(1,2)
  for(i=0; i<K; i++)
    for(j=0; j<K; j++)
      for(tr=0; tr<rloops; tr++)
        for(tc=0; tc<cloops; tc++)
#pragma PIPELINE
          for(tm=0; tm<Tm; tm++)
            for(tn=0; tn<Tn; tn++)
              if(i*j==0&&n==0)
                out[tm][tr][tc]=bias[tm]
              else
                wx=weights[tn][tm][i][j]
                ix=in[tn][S*tr+i][S*tc+j]
                out[tm][tr][tc]+=wx*ix

write_output():
  WR=ceil(Tm/MP)
  for(wr=0; n+1==nsteps&&wr<WR; wr++)
    for(p=0; p<MP; p++)
#pragma UNROLL, LOOP_MERGE
      xfer(out[WR*p+wr])

TOP():
  transfer_argument_descriptor()
  for(r=0; r<rsteps; r++)
    for(c=0; c<csteps; c++)
      for(m=0; m<msteps; m++)
        for(n=0; n<nsteps; n++)
#pragma DATAFLOW
          read_bias()    // omitted for brevity
          read_input()   // omitted for brevity
          read_weights() // omitted for brevity
          compute()
          write_output()
\end{lstlisting}

Each iteration of the top-level loops performs computation for one input tile.
The {\it DATAFLOW} directive ensures that the operations inside the $n$ loop are pipelined using ping-pong buffers for the concurrently accessed data structures.
The $in$ feature maps and $weights$ are read on every iteration, using the ping-pong buffers to allow reading the subsequent iteration's data while computation is in progress.
The $out$ feature maps are double buffered to allow the start of the subsequent computation while the write of the previous output is in progress; however, the $n+1=n_{steps}$ condition prevents transfer on all but the last input tile.
Bias read is similarly limited to occur only on the initial iterations to avoid unnecessary transfers.

To minimize port idle time, all transfer functions perform the maximum-length contiguous bursts allowed by the data structures.
To minimize the number of bursts performed, we prioritize the $n$ dimension over the $m$ dimension of the $weights$ array, as the CLP designs have $T_n$ smaller than $T_m$.
The \textit{read\_input()}, \textit{read\_weights()}, and \textit{write\_output()} functions are parameterized to support concurrent transfers across multiple ports by partitioning the transfers according to the top array dimension as demonstrated in \textit{write\_output()}.

The {\it PIPELINE} directive in \textit{compute()} unrolls the $T_m$ and $T_n$ loops to create the CLP compute module.
To ensure concurrent access to the data, the $out$, $in$, $bias$, and $weights$ arrays are partitioned across different memory banks.
The $K \times K$ loops are the outer-most dimension to avoid back-to-back (``loop-carry'') data dependencies across loop iterations.
$r_{loops}$ and $c_{loops}$ equal $T_r$ and $T_c$, except for the last iteration of the $r$ and $c$ loops, in which case $r_{loops}$ and $c_{loops}$ take the boundary into account.

\section{Evaluation}
\label{sec:evaluation}

\newcommand{\taba}{
\begin{tabular}[b]{@{}lrrrrrr@{}}
 \toprule
            & $T_n$ & $T_m$ & Layers & $T_r$ & $T_c$  & Cycles$(\times 1000)$\\
 \midrule
 CLP0       & 7 & 64 & 1a, 1b & 8  & 8  & 732\\
            &   &    & 2a, 2b & 14 & 27 & 510\\
            &   &    & 3a, 3b & 13 & 13 & 338\\
            &   &    & 4a, 4b & 13 & 13 & 256\\
            &   &    & 5a, 5b & 13 & 13 & 170\\
\midrule
Overall     &   &    &       &    &     & 2,006\\
 \bottomrule
\end{tabular}
}

\newcommand{\tabb}{
\begin{tabular}[t]{@{}lrrrrrr@{}}
 \toprule
            & $T_n$ & $T_m$ & Layers & $T_r$ & $T_c$ & Cycles$(\times 1000)$ \\
 \midrule
 CLP0       & 2 & 64 & 5a, 5b & 13 & 13 & 584\\
            &   &    & 4a, 4b & 13 & 13 & 876\\
 \midrule
 CLP1       & 1 & 96 & 3a, 3b & 13 & 13 & 1,558\\
 \midrule
 CLP2       & 3 & 24 & 1a, 1b & 14 & 19 & 1,464\\
 \midrule
 CLP3       & 8 & 19 & 2a, 2b & 14 & 27 & 1,531\\
\midrule
Overall     &   &    &       &    &    & 1,558\\
\bottomrule
\end{tabular}
}

\newcommand{\tabc}{
\begin{tabular}[b]{@{}lrrrrrr@{}}
 \toprule
            & $T_n$ & $T_m$ & Layers & $T_r$ & $T_c$ & Cycles$(\times 1000)$\\
 \midrule
 CLP0       & 9 & 64 & 1a, 1b & 8  & 8  & 732\\
            &   &    & 2a, 2b & 14 & 27 & 437\\
            &   &    & 3a, 3b & 13 & 13 & 265\\
            &   &    & 4a, 4b & 13 & 13 & 201\\
            &   &    & 5a, 5b & 13 & 13 & 134\\
\midrule
Overall     &   &   &       &    &    &  1,769\\
\bottomrule
\end{tabular}
}

\newcommand{\tabd}{
\begin{tabular}[t]{@{}lrrrrrr@{}}
 \toprule
            & $T_n$ & $T_m$ & Layers & $T_r$ & $T_c$ & Cycles$(\times 1000)$\\
 \midrule
 CLP0       & 1 & 64  & 5a, 5b & 13 & 13 & 1,168\\
 \midrule
 CLP1       & 1 & 96  & 4a, 4b & 13 & 13 & 1,168\\
 \midrule
 CLP2       & 2 & 64  & 3a, 3b & 13 & 13 & 1,168\\
 \midrule
 CLP3       & 1  & 48 & 1a   &  14 & 19 & 1,098\\
  \midrule
 CLP4       & 1  & 48 & 1b   &  14 & 14 & 1,098\\
   \midrule
 CLP5       & 3  & 64 & 2a, 2b &  27 & 27 & 1,166\\
\midrule
Overall     &    &    &      &    &    & 1,168\\
 \bottomrule
\end{tabular}
}

\newcommand{\sqtaba}{
\begin{tabular}[b]{@{}lrrrr@{}}
 \toprule
            & $T_n$ & $T_m$ & Layers & Cycles$(\times 1000)$\\
 \midrule
 CLP0       & 32 & 68 & 1--26 & 349\\
\midrule
Overall     &    &    &      & 349\\
 \bottomrule
\end{tabular}
}

\newcommand{\sqtabb}{
\begin{tabular}[t]{@{}lrrrr@{}}
 \toprule
            & $T_n$ & $T_m$ & Layers & Cycles$(\times 1000)$ \\
 \midrule
 CLP0        & 6 & 16 & 2,3,6,5      & 179\\
 \midrule
 CLP1        & 3 & 64 & 1,8,9,12     & 183\\
 \midrule
 CLP2        & 4 & 64 & all others   & 165\\
\midrule
 CLP3        & 8 & 64 & 7,4,16,19    & 176\\
\midrule
 CLP4        & 8 & 128 & 26,22,25,13 & 185\\
\midrule
 CLP5        & 16 & 10 & 10          & 183\\
\midrule
Overall      &    &    &             & 185\\
\bottomrule
\end{tabular}
}

\newcommand{\sqtabc}{
\begin{tabular}[b]{@{}lrrrr@{}}
 \toprule
             & $T_n$ & $T_m$ & Layers & Cycles$(\times 1000)$\\
 \midrule
 CLP0        & 32 & 87 & 1--26 & 331\\
\midrule
 Overall     &    &    &       & 331\\
\bottomrule
\end{tabular}
}

\newcommand{\sqtabd}{
\begin{tabular}[t]{@{}lrrrr@{}}
 \toprule
            & $T_n$ & $T_m$ & Layers & Cycles$(\times 1000)$ \\
 \midrule
 CLP0       & 8 & 16    & 2,6,3,5    & 125\\
 \midrule
 CLP1       & 3 & 64    & 1          & 115\\
 \midrule
 CLP2       & 11 & 32   & all others & 133\\
\midrule
 CLP3       & 8  & 64   & 7,4,16     & 145\\
\midrule
 CLP4       & 5 & 256   & 19,26,22,25 & 144\\
\midrule
 CLP5       & 16 & 26   & 13,10      & 141\\
\midrule
Overall     &   &    &              & 145\\
\bottomrule
\end{tabular}
}

We evaluate the Multi-CLP approach to CNN accelerator design by applying our method to four networks (AlexNet, VGGNet-E, SqueezeNet and GoogLeNet), targeting two Xilinx Virtex-7 FPGAs (485T and 690T).   We consider designs with both single precision floating-point and 16-bit fixed point arithmetic.

We use our optimization method (Section~\ref{sec:new-design}) to determine the best Single-CLP and Multi-CLP designs for each chip's available resources and compare the model results. Further, we implement the designs using the HLS-based method (Section~\ref{sec:hls}) and use simulation, synthesis, and place and route tools to compare the Single-CLP and Multi-CLP methods, and to quantitatively evaluate the correctness of our models.
To fairly compare with prior work, we first demonstrate that our Single-CLP design for AlexNet on a Virtex-7 485T FPGA with single precision floating point is equivalent to the design in~\cite{zhang2015optimizing}.

Overall, our results show that our Multi-CLP methodology yields improvements ranging from 1.01x (VGGNet-E on 485T, single precision floating point) to 3.8x (AlexNet on 690T, fixed point) compared to Single-CLP designs.

\subsection{Methodology}

We use the optimization procedure from Section~\ref{ssec:optimization} to find the highest-throughput Single-CLP and Multi-CLP designs for all designs. Optimization times range from less than a minute to less than an hour on one CPU.
As input to the optimization procedure, we set the DSP and BRAM targets to 80\% of the FPGA's capacity:
1,648 BRAMs and 2,240 DSP slices on the 485T, and 2,352 BRAMs and 2,880 DSP slices on the 690T.

\subsection{Utilization Benefits of Multi-CLP}

We first compare the dynamic arithmetic unit utilization of Single-CLP and Multi-CLP designs across the 16 cases (four networks, two data types, two FPGAs). For this comparison, we do not restrict bandwidth, examining the effectiveness of Multi-CLP in improving dynamic arithmetic unit utilization.

Table~\ref{table:utilizations} shows the arithmetic unit utilization of all designs. Multi-CLP achieves higher dynamic utilization than Single-CLP in all cases. The smallest improvement (1.01x) is seen when targeting VGGNet-E because the convolutional layers of VGGNet-E have very regular dimensions. The best improvement (3.8x) is seen when targeting AlexNet because the first layer of AlexNet requires a large amount of computation, but has a small $(N,M)$ of $(3,48)$. Multi-CLP gives significant improvements on SqueezeNet (2.1x) and GoogLeNet (2.0x), showing that it benefits a wide range of convolutional neural networks, including large and deep networks like GoogLeNet. Also noteworthy is that larger improvements are seen when the number of available arithmetic units increases---both on the larger FPGA (690T) and when using fixed-point arithmetic (meaning more arithmetic is possible using the same resources). This demonstrates that Single-CLP has an inherent scaling problem: as the number of arithmetic units increases, a Single-CLP struggles to use them all. Conversely, our Multi-CLP design makes much more efficient use of the available units.

\begin{table}
\caption{Dynamic arithmetic unit utilization of the Single-CLP designs and Multi-CLP designs.}
\label{table:utilizations}
\centering
\tabcolsep=0.11cm
\begin{tabular}{@{}lrrrr@{}}
\toprule
                      & AlexNet & VGGNet-E  & SqueezeNet & GoogLeNet\\
\midrule
{\bf 485T (float)}\\
S-CLP                 & 74.1\%     & 96.8\% & 78.0\% & 81.9\%\\
M-CLP                 & 95.4\%     & 97.5\% & 95.8\% & 96.9\%\\
\midrule
{\bf 690T (float)}\\
S-CLP                 & 65.4\%     & 96.0\% & 76.4\% & 78.1\%\\
M-CLP                 & 99.0\%     & 98.7\% & 96.7\% & 96.0\%\\
\midrule
{\bf 485T (fixed)}\\
S-CLP                 & 31.0\%     & 89.7\% & 51.1\% & 50.2\%\\
M-CLP                 & 93.9\%     & 97.3\% & 93.6\% & 93.8\%\\
\midrule
{\bf 690T (fixed)}\\
S-CLP                 & 23.7\%     & 88.3\% & 42.0\% & 44.0\%\\
M-CLP                 & 90.6\%     & 96.1\% & 93.1\% & 89.3\%\\
\bottomrule
\end{tabular}
\end{table}

\subsection{Detailed Comparison: Single- vs Multi-CLP}

To examine the differences between Single-CLP and Multi-CLP designs, we present detailed comparisons for two networks and scenarios.
First, to compare with the Single-CLP design in~\cite{zhang2015optimizing}, we choose the same network and parameters: AlexNet using floating point at 100MHz.
Then, we evaluate a more aggressive scenario: SqueezeNet using 16-bit fixed-point arithmetic at 170MHz.

Tables~\ref{tab:alex-float-config} and~\ref{tab:squeezenet-fixed-config} present the parameters chosen by our optimization for AlexNet and SqueezeNet on each FPGA.
In each table, $T_n$ and $T_m$ give the parallelism of the compute module (Figure~\ref{fig:clp}).
Additionally, for AlexNet we show the $T_r$ and $T_c$ parameters, which control the on-chip data tiling (Section~\ref{ssec:clp}).

\begin{table*}
\centering
\caption{AlexNet, 32-bit floating point: Single-CLP and Multi-CLP accelerators.}
\subfloat[485T Single-CLP]{\taba}\quad\quad
\subfloat[690T Single-CLP]{\tabc}

\subfloat[485T Multi-CLP]{\tabb}\quad\quad
\subfloat[690T Multi-CLP]{\tabd}

\label{tab:alex-float-config}
\end{table*}

For AlexNet, Table~\ref{tab:alex-float-config} shows that when we target the same system as~\cite{zhang2015optimizing} (Single-CLP, 32-bit floating point, 485T), our optimization yields the same parameters
($T_n=7$ and $T_m=64$) and the same speed (2.0~million cycles).\footnote{The cycle counts in~\cite{zhang2015optimizing} only account for half of the convolutional layers (i.e., layers 1a, 2a, ..., 5a of Figure~\ref{fig:alexnet}, but not layers 1b, 2b, ..., 5b). We therefore double the cycle count in Table 4 of~\cite{zhang2015optimizing} to compare with our implementation.}\textsuperscript{,}\footnote{Prior work~\cite{zhang2015optimizing} does not report $T_r$ and $T_c$.}
Accenting the fairness of the comparison, we note that the Single-CLP and Multi-CLP designs have the same arithmetic unit cost, which the Multi-CLP design spreads among several CLPs.  Recall that a CLP requires $T_n\times T_m$ multipliers and adders.
For example, on the 690T FPGA, the Single-CLP design uses $9\times64=576$ multipliers and adders in one CLP.
The corresponding Multi-CLP design uses the same number ($1\times64 + 1\times96 + 2 \times 64 + 1 \times 48 + 1 \times 48 + 3 \times 64=576$), but distributes them over six CLPs.

Table~\ref{tab:alex-float-config} also shows which of the 10 convolutional layers of the network map to which CLP (e.g., for the Multi-CLP on the 485T, CLP0 executes layers 4a, 4b, 5a, and 5b).
The last column of each table shows the number of cycles (in terms of cycles $\times 1000$) that each CLP takes to execute its layers, with the overall cycles per image for each system shown at the bottom.
For the Single-CLP designs, the overall cycle count is the sum of how long the CLP takes to compute all ten layers.
When multiple layers are listed in the same row (such as 4a and 4b), the cycle count is the number of cycles needed to execute all of those layers.

In the Multi-CLP system, the CLPs operate concurrently. The overall cycle count for the accelerator is the maximum of the cycle counts of its CLPs,
because this dictates the epoch length (the interval between times when the pipelined Multi-CLP system is able to start processing a new image).
For example, in the AlexNet 485T Multi-CLP design, the four CLPs have cycle counts of $584+876=1460$, $1558$, $1464$, and $1531$ thousand cycles, giving an overall time of $1558$ thousand cycles.

Because our optimization maximizes the overall throughput, the Multi-CLP designs it produces tend to be balanced.
This balance indicates that the resources are effectively spread across the computation pipeline, such that each CLP can be kept busy most of the time.
Table~\ref{table:performance} shows the arithmetic unit utilization of each AlexNet design, as well as the throughput (for convolutional layers) and the modeled consumption of DSP slices, BRAMs, and bandwidth.
We see that on both FPGAs, the Multi-CLP designs provide a significant throughput
advantage over the Single-CLP: 1.31x on the 485T and 1.54x on the 690T.
Because the Single- and Multi-CLP designs use an equal number of arithmetic units (built from the same number of DSP slices), the speedup is proportional to the Multi-CLP improvement in arithmetic unit utilization.
The 485T and 690T Single-CLP designs are only able to provide useful work to the arithmetic units 72.6\% and 64.0\% of the time, respectively, while Multi-CLP improves utilization to 95.1\% and 98.9\%.
The GFlop/s rate (in the last column) is proportional to the throughput.

\begin{table}
\caption{AlexNet, floating point: Model-predicted resource usage and throughput.
Bandwidth and throughput at 100MHz.}
\label{table:performance}
\centering
\tabcolsep=0.11cm
\begin{tabular}{@{}lrrrrrr@{}}
\toprule
                      & BRAM & DSP  & \shortstack{B/w\\(GB/s)} & \shortstack{Arith\\Util.(\%)} & \shortstack{Thr.\\(img/s)} & Gflop/s\\
\midrule
{\bf 485T}\\
S-CLP              & 618          & 2,240 & 1.40               &         72.6  & 48.85 & 65.05\\
M-CLP              & 731          & 2,240 & 1.38               &         95.1  & 63.98 & 85.20\\
\midrule
{\bf 690T}\\
S-CLP              & 758          & 2,880 & 1.78               &         64.0 & 55.40 & 73.77\\
M-CLP              & 1,238        & 2,880 & 1.49               &         98.9 & 85.55 & 113.92\\
\bottomrule
\end{tabular}
\end{table}

As the rate of computation increases, the amount of data that must be kept on chip increase commensurately.
On the 485T FPGA, the 1.31x throughput improvement comes at a cost of 1.18x higher BRAM usage.
On the 690T, the throughput improvement of the Multi-CLP designs grows to 1.54x, requiring 1.63x higher BRAM usage.
However, it is worth noting that there is a tradeoff between the number of BRAMs used and off-chip memory bandwidth.
We can save bandwidth by adding more input and output buffers, or we can reduce buffer sizes at the cost of higher bandwidth.

\begin{figure}
\centering
\includegraphics[width=\columnwidth]{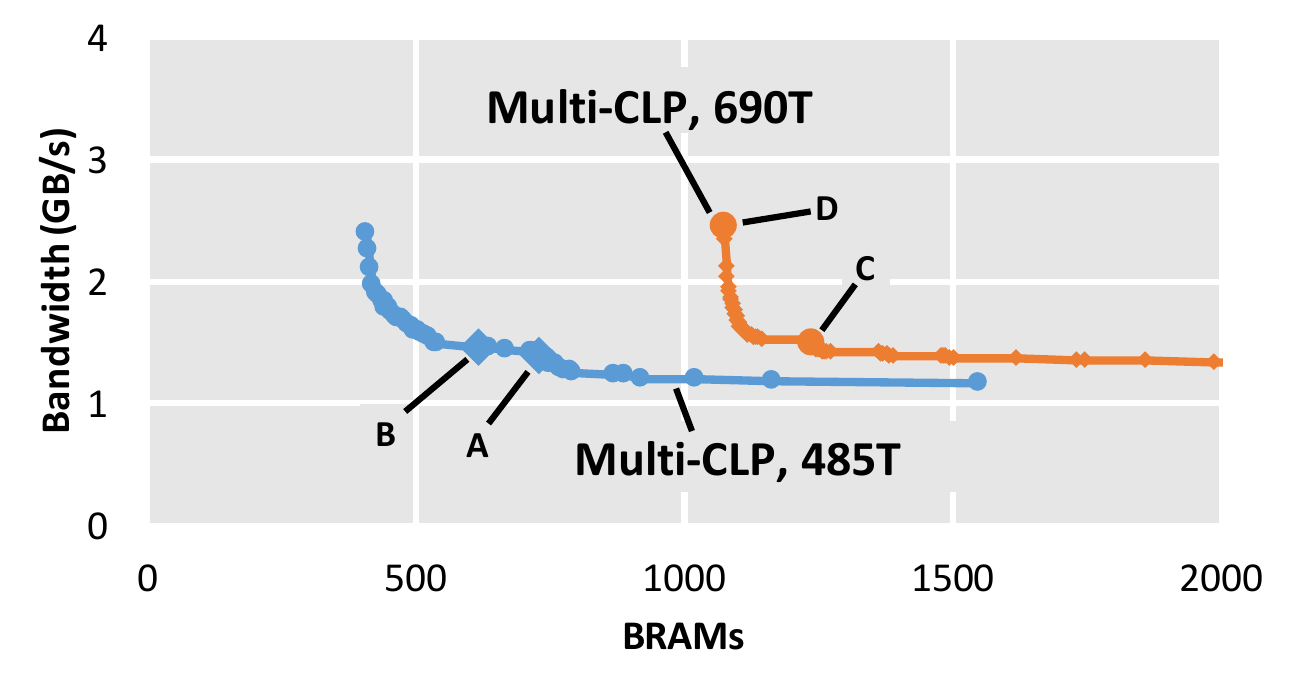}
\caption{Tradeoff between BRAM usage and off-chip memory bandwidth required for Multi-CLP designs.}
\label{fig:bram-vs-bandwidth}
\end{figure}

We illustrate this phenomenon in Figure~\ref{fig:bram-vs-bandwidth}, showing the options for the two Multi-CLP designs.
The Multi-CLP designs shown in Table~\ref{table:performance} were chosen to roughly match the memory bandwidth used by the Single-CLP system.
However, one could also adjust this tradeoff, saving BRAMs while using more bandwidth.
All alternatives for each system have nearly identical throughput (e.g., all 690T designs have the same throughput as shown in the table, with the differences bounded by $step$ in Listing~\ref{list:optimize}), but each makes a different tradeoff between BRAM capacity and off-chip bandwidth.
For example, the points labeled A and C correspond to the iso-bandwidth designs described above.
Another useful example is represented by points B and D, which allow the Multi-CLP designs to approximate the BRAM utilization of the Single-CLP designs, at the expense of bandwidth.
Point B reduces the 485T Multi-CLP BRAM usage to 619, but increases the bandwidth requirement to 1.46 GB/s.
On the 690T FPGA, point D represents an alternate design using only 1075 BRAMs, but requiring a bandwidth of 2.44 GB/s.
Given specific bandwidth and BRAM constraints, the optimization tool or the designer can choose between different points along the curve.

Tables~\ref{tab:squeezenet-fixed-config} and~\ref{table:squeezenet-performance} present the 16-bit fixed-point SqueezeNet designs at 170MHz.
Here, we expect the accelerator to be bandwidth bound, so we direct the optimizer to use a heuristic that groups layers by their compute-to-data ratios (Section~\ref{ssec:optimization}).
To reduce optimization time, we limit the number of CLPs to at most six.
Similar to the AlexNet case, each CLP of the Multi-CLP SqueezeNet accelerator finishes its work in roughly the same time, minimizing idling due
to work imbalance. The results show a dramatic improvement in arithmetic utilization and thus throughput---up to a 2.33x improvement over the Single-CLP design on the same FPGA.
We also observe that the peak bandwidth for SqueezeNet is significantly higher than AlexNet---due to the characteristics of the network and because the accelerators run at a higher clock frequency.
Although the two Multi-CLP SqueezeNet designs require 1.23x and 1.32x more BRAMs than the Single-CLP designs, they provide 1.91x and 2.33x higher throughput with a lower off-chip bandwidth requirement.

To estimate the bandwidth required for a CNN to reach peak throughput on a given FPGA, we set $Cycles_{min}$ in Listing~\ref{list:optimize} to match the best performance when bandwidth is unlimited, then we gradually relax the bandwidth constraint $bw$ until solutions within 2\% of $Cycles_{min}$ can be found.
The 2\% margin is set to avoid overestimating bandwidth requirements due to instantaneous data transfer spikes.
Throughputs in Tables~\ref{table:performance} and~\ref{table:squeezenet-performance} are bandwidth-optimized, whereas cycle counts in Tables~\ref{tab:alex-float-config} and~\ref{tab:squeezenet-fixed-config} are bandwidth-unconstrained.

\begin{table*}
\centering
\caption{SqueezeNet, 16-bit fixed-point: Single-CLP and Multi-CLP accelerators.}
\subfloat[485T Single-CLP]{\sqtaba}\quad\quad
\subfloat[690T Single-CLP]{\sqtabc}

\subfloat[485T Multi-CLP]{\sqtabb}\quad\quad
\subfloat[690T Multi-CLP]{\sqtabd}

\label{tab:squeezenet-fixed-config}
\end{table*}

\begin{table}
\caption{SqueezeNet, fixed point: Model-predicted resource usage and throughput.
Bandwidth and throughput at 170MHz.}
\label{table:squeezenet-performance}
\centering
\tabcolsep=0.11cm
\begin{tabular}{@{}lrrrrrr@{}}
\toprule
                      & BRAM & DSP  & \shortstack{B/w\\(GB/s)} & \shortstack{Arith\\Util.(\%)} & \shortstack{Thr.\\(img/s)} & Gop/s\\
\midrule
{\bf 485T}\\
S-CLP            & 400          & 2,176 & 19.7               &         50.3 & 480.0 & 372.2 \\
M-CLP            & 492          & 2,240 & 15.3               &         93.0 & 913.4 & 708.3 \\
\midrule
{\bf 690T}\\
S-CLP            & 480          & 2,784 & 20.5               &         41.3 & 504.1 & 391.0 \\
M-CLP            & 635          & 2,880 & 19.5               &         92.9 & 1173.0 & 909.7\\
\bottomrule
\end{tabular}
\end{table}

\subsection{Model Validation}

To validate our model, we synthesized and implemented (place and route) four designs using the HLS-based template described in Section~\ref{sec:hls}.
We place-and-route the CLPs in isolation, omitting infrastructure like AXI crossbars and memory controllers.
First, we use our methodology to design a 32-bit floating point Single-CLP for the 485T FPGA running at a 100MHz---this enables a direct comparison to the Single-CLP HLS results in~\cite{zhang2015optimizing}.
Then, we evaluate Multi-CLP designs for AlexNet on the 485T and 690T FPGAs, and a Multi-CLP design for SqueezeNet on the 690T.
Tables~\ref{tab:model-vs-impl} (AlexNet) and~\ref{tab:model-vs-impl-squeezenet} (SqueezeNet) compare our model predictions with the implementation results in terms of DSP slices and BRAMs.
For the Multi-CLP designs, we compare the metrics of each CLP in addition to the overall values for the entire Multi-CLP design.

\begin{table}
\caption{AlexNet, 32-bit floating point: comparison of model prediction and implementation results.}
\label{tab:model-vs-impl}
\centering
\tabcolsep=0.13cm
\begin{tabular}{@{}lrrcrr@{}}
 \toprule
& \multicolumn{2}{c}{BRAM} &~& \multicolumn{2}{c}{DSP} \\
\cmidrule{2-3}\cmidrule{5-6}
             & model & impl. && model & impl.   \\
\midrule
\multicolumn{4}{@{}l}{\bf 485T Single-CLP}\\
CLP0       & 618    & 698     && 2,240   & 2,309     \\
\midrule
\multicolumn{4}{@{}l}{\bf 485T Multi-CLP}\\
CLP0       & 130    & 132   && 640    & 689   \\
CLP1       & 193    & 195   && 480    & 529   \\
CLP2       & 186    & 242   && 360    & 410   \\
CLP3       & 222    & 243   && 760    & 815   \\
Overall    & 731    & 812   && 2,240  & 2,443   \\
\midrule
\multicolumn{4}{@{}l}{\bf 690T Multi-CLP}\\
CLP0       & 129    & 131     && 320    & 369   \\
CLP1       & 193    & 195     && 480    & 529   \\
CLP2       & 130    & 132     && 640    & 689   \\
CLP3       & 166    & 226     && 240    & 290  \\
CLP4       & 160    & 162     && 240    & 290  \\
CLP5       & 460    & 590     && 960    & 1,010  \\
Overall    & 1,238  & 1,436   && 2,880  & 3,177 \\
\bottomrule
\end{tabular}
\end{table}

\begin{table}
\caption{SqueezeNet 16-bit fixed point: comparison of model prediction and implementation results.}
\label{tab:model-vs-impl-squeezenet}
\centering
\tabcolsep=0.13cm
\begin{tabular}{@{}lrrcrr@{}}
 \toprule
& \multicolumn{2}{c}{BRAM} &~& \multicolumn{2}{c}{DSP} \\
\cmidrule{2-3}\cmidrule{5-6}
             & model & impl. && model & impl.   \\
\midrule
\multicolumn{4}{@{}l}{\bf 690T Multi-CLP}\\
CLP0       & 24     & 42     && 128    & 227    \\
CLP1       & 152    & 218     && 192    & 264    \\
CLP2       & 44     & 78     && 352    & 508   \\
CLP3       & 72     & 138     && 512    & 592  \\
CLP4       & 259    & 520     && 1,280   & 1,416   \\
CLP5       & 84     & 112     && 416    & 478  \\
\midrule
Overall    & 635    & 1,108   && 2,880    & 3,494 \\
\bottomrule
\end{tabular}
\end{table}

The model predictions are extremely close to the implemented system,
with only minor divergences.
For example, the model underestimates the DSP counts by approximately 50--100 DSP slices per CLP, as the model accounts only for the DSP slices used for the compute module of the convolution arithmetic and does not include DSP slices that are used in the address calculations, loop indexing, and control logic.
By examining the resource utilization of specific components, we verified that the number of DSP slices used in the compute modules perfectly matches the model prediction.
Similarly, we find small discrepancies between the predicted and implemented BRAM counts, caused by the way the tools map memories.

We conclude that the differences observed between the model and the implementation results are not symptoms of the model not matching the design and its requirements. Instead, the differences occur because the model does not take into account some toolflow-specific considerations.
Several minor modifications could be made to correct these errors, at the cost of making the model toolflow-specific.
Furthermore, we also performed RTL simulation of the resulting designs; the simulated number of cycles only differs from our model by the pipeline depth of the implementation.

\subsection{CNN Accelerator Resource Utilization}

Tables~\ref{table:resource} and~\ref{table:resource-squeezenet} report the total resources (including FPGA flip-flops and LUTs) used by each of the designs, extracted after place-and-route.
The counts include the CLPs only, not including platform-specific memory controllers or crossbars.
Closely following our model validation results, we observe that, for the same FPGA target, a Multi-CLP implementation uses more DSP slices than the corresponding Single-CLP implementation.
Although the compute modules (i.e., the arithmetic units used for the convolution's multiplications and additions) use the same number of DSP slices, a difference arises due to the logic for address calculation and loop indexing, adding approximately 6\% DSP slices to the Multi-CLPs designs.
Similar increases are seen in the flip-flop and LUT counts; more CLPs require additional control logic beyond the DSP slices and BRAMs.
However, ultimately, the DSP slices limit the implementations significantly more than the flip-flops or LUTs.
For completeness, we use Vivado to produce post-place-and-route power estimates, which are reported in Watts for each design.

\begin{table}
\caption{AlexNet, 32-bit floating point: FPGA resource utilization and estimated power for the Single-CLP and Multi-CLP designs optimized for the 485T and 690T.}
\label{table:resource}
\centering
\begin{tabular}{@{}lrrcr@{}}
\toprule
& \multicolumn{2}{c}{485T} &&\multicolumn{1}{c}{690T}\\
\cmidrule{2-3}\cmidrule{5-5}
            & Single-CLP   & Multi-CLP   &&  Multi-CLP\\
\midrule
BRAM-18K    & 698     & 812     && 1,436\\
            & (34\%)  & (39\%)  && (49\%)\\
\midrule
DSP         & 2,309   & 2,443   && 3,177 \\
            & (82\%)  & (87\%)  && (88\%)\\
\midrule
FF          & 219,815 & 270,991 && 348,049 \\
            & (36\%)  & (45\%)  &&  (40\%)\\
\midrule
LUT         & 146,325 & 176,876 &&  236,877 \\
            & (48\%)  & (58\%)  && (55\%)\\
\midrule
Power       & 6.6 W   & 7.6 W   && 10.2 W\\
\bottomrule
\end{tabular}
\end{table}

\begin{table}
\caption{SqueezeNet, 16-bit: FPGA resource utilization and estimated power for a
Multi-CLP system optimized for the 690T.}
\label{table:resource-squeezenet}
\centering
\begin{tabular}{@{}rrrrr@{}}
\toprule
BRAM-18K & DSP & FF & LUT & Power\\
\midrule
1,108  &~~~ 3,494  &~~~ 161,411 &~~~ 133,854  &~~~ 7.2 W \\
(38\%)   & (97\%)   & (19\%)  & (31\%) \\
\bottomrule
\end{tabular}
\end{table}

\subsection{Projections to Future FPGAs}
\label{ssec:future_fpgas}

\begin{figure}[t]
  \centering
  \includegraphics[width=\columnwidth]{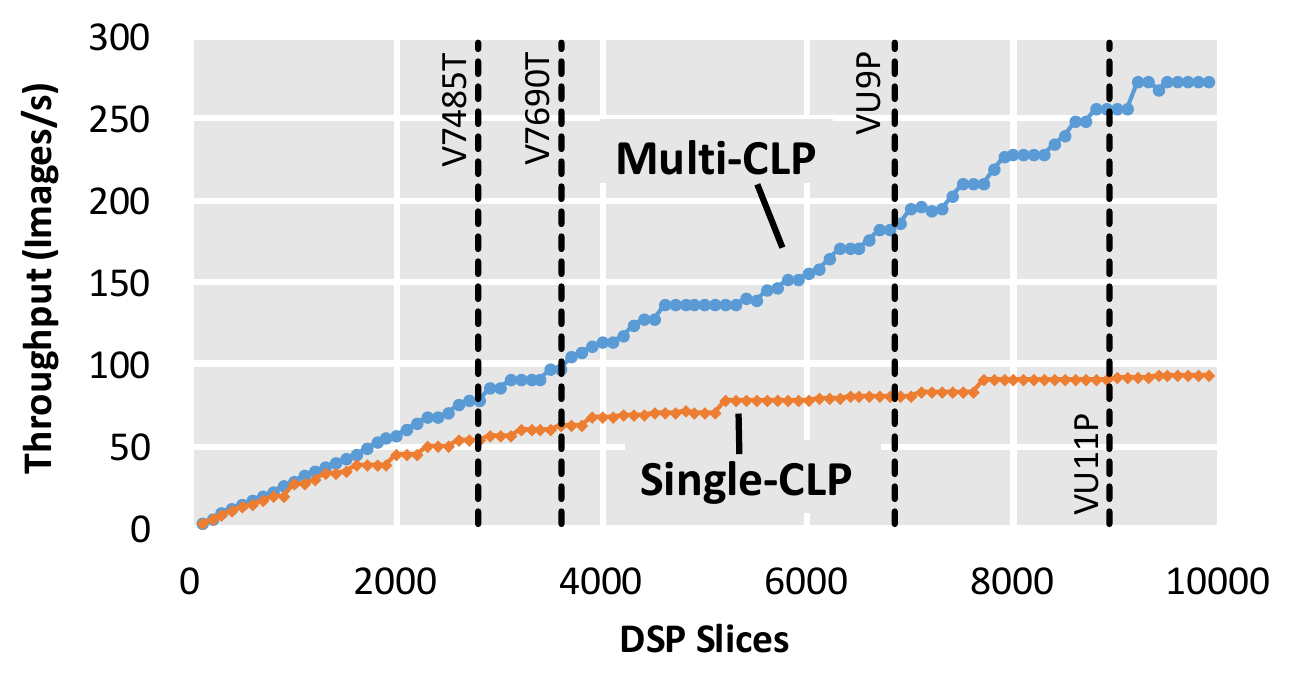}
\caption{Throughput at 100MHz for AlexNet on Multi-CLP and Single-CLP 32-bit floating point designs as a function of the number of available DSP slices.}
  \label{fig:sweep_figure}
\end{figure}

In addition to comparing the Single-CLP and Multi-CLP designs on the Virtex-7 FPGAs, it is also instructive to examine how the Single-CLP and Multi-CLP designs scale as the FPGA resource budget grows.
For example, the Xilinx roadmap includes UltraScale+ devices with over 10,000 DSP slices.
Figure~\ref{fig:sweep_figure} projects the throughput of the Multi-CLP and Single-CLP floating point AlexNet designs for DSP slice budgets ranging from 100 to 10,000.
For each point, we perform an optimization to find the best Single-CLP and Multi-CLP designs and report the estimated throughput.

The x-axis shows the number of DSP slices used for each design.
The BRAM budget is set at a ratio of one BRAM (18Kb) to every 1.3 DSP slices, an approximation of the relationship we observe in the Virtex-7 parts.
Dashed vertical lines illustrate the total number of DSP slices available on the Virtex-7 485T, Virtex-7 690T, Virtex UltraScale+ 9P, and Virtex UltraScale+ 11P FPGAs.
Note that the dashed lines are provided only for the perspective of resource capacity; real designs constrain the resources available to the CNN accelerator below the full chip capacity.

As the number of available DSP slices increases, the throughput difference between the Single- and Multi-CLP designs grows.
For example, going from 2,240 to 9,600 DSP slices, the Multi-CLP improvement over Single-CLP designs increases from 1.3x to 3.3x.

\section{Related Work}
\label{sec:related}
    Eyeriss~\cite{chen2016eyeriss_jssc, chen2016eyeriss_isca} is a recent ASIC CNN accelerator that couples a compute grid with a NoC, enabling flexibility in scheduling CNN computation. This flexibility limits arithmetic unit underutilization. However, underutilization still exists when a CNN layer's kernel size and output feature map height are incompatible with the dimensions of the compute grid.

    \cite{li2016high} proposes an FPGA accelerator for AlexNet that has one module per layer, and thus can achieve high arithmetic unit utilization.
    However, this design stores all intermediate data of the network on chip, limiting the size of the network that can be supported with this approach. Moreover, as discussed in Section~\ref{ssec:vpipe}, building one module per layer does not work well for larger networks.

    Our baseline Single-CLP design is based on~\cite{zhang2015optimizing}. Similar designs are used in~\cite{chen2014diannao,chen2014dadiannao}.
    Other recent works propose different CNN acceleration hardware.
    For example,~\cite{farabet2009cnp,farabet2010hardware,farabet2011neuflow,sankaradas2009massively,chakradhar2010dynamically} focus on 2D-convolvers, which play the roles of both compute modules and data caches.
    Meanwhile,~\cite{peemen2013memory,peemen2015inter} use FMA units for computation.
    The key differences between these approaches are the order of data transfer and the choice of memory organization.
    Several key similarities cause these methods to suffer from the underutilization problem we observe in our Single-CLP design.
    For example, the 2D-convolvers used in~\cite{farabet2009cnp,farabet2010hardware,sankaradas2009massively,chakradhar2010dynamically} must be provisioned for the largest filter across layers; they will necessarily be underutilized when computing layers with smaller filters.
    In~\cite{peemen2013memory}, the organization of the compute modules depends on the number of output feature maps and their number of rows.
    Both of these parameters typically change across layers, resulting in an analogous resource underutilization problem.
    Our Multi-CLP resource partitioning technique can used by these designs to improve arithmetic unit utilization.

    C-Brain~\cite{song2016cbrain} offers an orthogonal approach, transforming a stride-$S$ convolution to $S$ stride-1 convolutions to increase PE utilization for CLPs. However, this method can only be used when the convolution stride of a layer is greater than one and the effectiveness of the technique depends on the stride size.

    Several recent works explored other promising, but orthogonal, aspects of CNN accelerators.
    \cite{albericio2016cnvlutin} proposes a CNN accelerator design that can skip computations on input values that are zeros.
    \cite{qiu2016going,judd2016proteus} reduce an accelerator's bandwidth and buffer use.
    \cite{qiu2016going} uses per-layer data quantization and matrix-decomposition, whereas \cite{judd2016proteus} uses per-layer numerical precision reduction.
    \cite{alwani2016fused} uses a fused-layer technique to reduce bandwidth use of convolutional layers.
    \cite{shen2017escher} optimizes batch sizes to reduce off-chip data transfer.
    These techniques can be integrated into Multi-CLP designs.

    \cite{sharma2016dnnweaver} and~\cite{wang2016deepburning} propose complete frameworks for generating FPGA-based accelerators from CNN specifications.
    Our Multi-CLP approach can be integrated into these frameworks to improve the performance of auto-generated accelerators.
    \cite{chi2016prime} and~\cite{shafiee2016isaac} explore in-memory-processing to accelerate CNNs.
    \cite{suda2016throughput} develop an OpenCL-based HLS tool to implement CNN accelerators that use different modules for different kinds of layers, but all convolutional layers are computed with a single CLP.

\section{Conclusions}
\label{sec:conclusions}
The traditional approach to FPGA-based CNN accelerator design follows a ``one size fits all'' methodology, where a single convolutional layer processor (CLP) computes all convolutional layers of the CNN.
In this paper, we observed that variation in the dimensions of the CNN layers limits the throughput of this ``Single-CLP'' approach; on layers whose dimensions are a poor fit for the CLP parameters, the arithmetic units exhibit low dynamic utilization, where adders and multipliers frequently remain idle.
To overcome this inefficiency, we presented a new design paradigm that partitions hardware resources among multiple cooperating CLPs.
Our ``Multi-CLP'' approach allows the CLP dimensions to more closely match the CNN layer dimensions, resulting in better dynamic resource utilization and higher throughput.

The optimization algorithm we developed finds efficient Multi-CLP designs within a given resource budget (DSP slices, BRAMs, and bandwidth).
For example, on the Virtex-7 690T FPGA, we showed that a Multi-CLP accelerator yields a 3.8x higher throughput than the state-of-the-art Single-CLP design, when accelerating AlexNet with 16-bit fixed point arithmetic, corresponding to an improvement of dynamic utilization from 24\% to 91\%. For the more recent SqueezeNet and GoogLeNet networks, our method results in speedups of 2.2x and 2.0x, respectively.
Further, we showed that the disparity between the throughput of the Single-CLP and Multi-CLP designs grows rapidly as the resource budget increases.

\begin{acks}
The authors would like to thank Cheng-Yang Fu and Alex C. Berg from the Computer Vision Group at the University of North Carolina at Chapel Hill for their help.
This material is based on work supported by the National Science Foundation (NSF) under Grant Nos. 1533739 and 1453460.
The experiments were conducted with equipment purchased through NSF CISE Research Infrastructure Grant No. 1405641.
\end{acks}

\bibliographystyle{ACM-Reference-Format}
\bibliography{refs}

\end{document}